\documentclass[conference,a4paper]{IEEEtran}
\IEEEoverridecommandlockouts
\usepackage{cite}
\usepackage{amssymb,amsfonts}
\usepackage{graphicx}
\usepackage{textcomp}
\usepackage{xcolor}
\usepackage{enumitem}
\usepackage{tikz}

\usepackage[cmex10]{amsmath}
\usepackage{array}

\usepackage{mathptmx,stmaryrd}

\newcommand{\eea}{\end{eqnarray}}
\newcommand{\bea}{\begin{eqnarray}}

\newcommand{\bali}{\begin{align}}

\newcommand{\bean}{\begin{eqnarray*}} 
\newcommand{\eean}{\end{eqnarray*}} 

\newcommand{\eeq}{\end{equation}}
\newcommand{\beq}{\begin{equation}}

\newcommand{\mcC}{\ensuremath{\mathcal{C}}}

\newcommand{\mcP}{\ensuremath{\mathcal{P}}}

\newcommand{\mcX}{\ensuremath{\mathcal{X}}}

\newcommand{\mcY}{\ensuremath{\mathcal{Y}}}
\newcommand{\mcW}{\ensuremath{\mathcal{W}}}

\newcommand{\lra}{\ensuremath{\leftrightarrow}}

\newcommand{\lp}{\ensuremath{\left (}}
\newcommand{\rp}{\ensuremath{\right )}}

\newcommand{\llb}{\ensuremath{\llbracket}} 
\newcommand{\rrb}{\ensuremath{\rrbracket}}

\usetikzlibrary{shapes,arrows,positioning}
\def\BibTeX{{\rm B\kern-.05em{\sc i\kern-.025em b}\kern-.08em
    T\kern-.1667em\lower.7ex\hbox{E}\kern-.125emX}}
\newtheorem{theorem}{Theorem}

\newtheorem{definition}{Definition}

\begin{document}

\title{Zero-Error Capacity of Multiple Access Channels via Nonstochastic Information\\
\thanks{This work was supported by the Australian Research Council under the Future Fellowships grant FT140100527.}
}

\author{\IEEEauthorblockN{Ghassen Zafzouf, Girish N.~Nair and Jamie S.~Evans}
\IEEEauthorblockA{\textit{Department of Electrical and Electronic Engineering}\\
\textit{University of Melbourne}\\
VIC 3010, Australia\\
gzafzouf@student.unimelb.edu.au, \{gnair, jse\}@unimelb.edu.au
}
}

\maketitle

\begin{abstract}
The problem of characterising the \emph{zero-error capacity region} for multiple access channels even in the noiseless case has remained an open problem for over three decades. Motivated by this challenging question, a recently developed theory of nonstochastic information is applied to characterise the zero-error capacity region for the case of two correlated transmitters.  
Unlike previous contributions, this analysis does not assume that the blocklength is asymptotically large. Finally, 
a new notion of nonstochastic information is proposed for a non-cooperative problem involving three agents. These results are preliminary steps towards understanding  information flows in worst-case distributed estimation and control problems. 
\end{abstract}

\section{Introduction}
The \emph{multiple access channel} (MAC) was initially introduced by Shannon in his work \cite{b3}. The multiple access communication system consists of several senders  that aim to transmit each an independent message reliably to a common receiver. This model corresponds indeed to various real-life scenarios such as multiple ground stations communicating with a satellite receiver, or the uplink phase of a cellular system. Clearly, the challenge in this case is not only the channel noise distorting the transmitted signal, but also the interference between the senders. The ordinary capacity region $\mathcal{C}$ of MAC channels has been extensively studied in the literature \cite{b4,b5,b6}, and by means of superposition coding, the single-letter characterization of this region was found by Slepian and Wolf \cite{b6}. It consists of the closure of the convex hull for all nonnegative rate tuples $(R_0,R_1,R_2)$ satisfying
\begin{align}
      R_1 & \leq I(X_1;Y \vert X_2,U),   \nonumber \\
      R_2 & \leq  I(X_2;Y \vert X_1,U),  \nonumber \\
      R_1 + R_2 & \leq  I(X_1, X_2;Y \vert U),  \nonumber \\
      R_0 + R_1 + R_2 & \leq  I(X_1, X_2;Y)
\label{slepian-wolf}
\end{align}
where $X_1 \leftrightarrow U \leftrightarrow X_2$ and $U \leftrightarrow X_1, X_2 \leftrightarrow Y $ form Markov chains. \\
A further important notion in addition to the ordinary capacity  is the so-called \emph{zero-error capacity}. This parameter is defined as the least upper bound of rates leading to an error probability at the receiver which is \emph{exactly} equal to zero \cite{b7}.   The significance of zero-error capacity has recently been shown in worst-case control problems where strict, deterministic guarantees on performance must be met \cite{matveevIJC07}.
However, little is  known about the zero-error capacity region of many simple MAC's. For instance, for deterministic binary adder channels, the best outer bound on this region has been found by Ordentlich and Shayevitz in \cite{b2} and presents a slight improvement on the result obtained by Urbanke and Li \cite{b8}. These studies mainly rely on combinatorics in order to tighten the outer bound of $\mathcal{C}_0$. 

In this paper we apply the concept of \emph{nonstochastic information} \cite{b1} to obtain an intrinsic  characterisation of the zero-error capacity region of a general noisy MAC. A motivation for investigating such a problem arises from the study of decentralised control systems. In fact, the independent senders model the sensors reading the states of different plants, while the common decoder can be seen as the controller stabilising the system. Furthermore, the concept of zero-error capacity has increasingly gained more attention as it is an insightful parameter of the system worst-case performance. Contrary to communication systems, in control applications safety presents a crucial criterion, and hence, the plant performance must be guaranteed not only on average but rather at all times. 
Thus, in this case $\mathcal{C}_0$ can be considered a more useful figure of merit than the classical Shannon capacity $\mathcal{C}$ which allows an arbitrary small probability of error. 

The rest of the paper is organised as follows. In Section \ref{sec:StationaryMemorylessUncertainMAC}, some basic definitions related to the nonprobabilistic framework are introduced and the  MAC model along with the zero-error coding scheme are presented.  Next, the zero-error capacity region for the MAC channel for any given block-length $n$ is characterised in Section \ref{sec:ZECapacity}, with converse and achievability proofs provided. A new notion of information in the MAC setting, namely the noncooperative $NC-$sense connectedness, is  studied in Section \ref{sec:NCConnectedness}. Finally, Section \ref{sec:conclusion} concludes the article by summarising the main contributions and discussing possible future directions.  

\section{Zero-Error Communication over MACs in the Nonstochastic Framework}
\label{sec:StationaryMemorylessUncertainMAC}

In this section, we reformulate the problem of zero-error communication over multiple access channels (MACs) in  terms of the nonstochastic framework of \cite{b1}. 

\subsection{Uncertain Variables, Unrelatedness and Markovianity}

First we briefly those elements of the nonstochastic framework of \cite{b9} that are needed for this section. We present further aspects as required in subsequent sections. 


An \emph{uncertain variable}  (uv) $X$ consists 
of a mapping from an underlying sample space $\Omega$ to a
space $\mathcal{X}$ of interest \cite{b9}. Each sample $\omega\in\Omega$ is hence mapped to a particular realization $X(\omega)\in \mathcal{X}$. 
For a pair of uv's $X$ and $Y$,  we denote their \emph{marginal}, \emph{joint} and \emph{conditional ranges} as 
\begin{align}
   [\![ X ]\!] & := \lbrace X(\omega): \omega \in \Omega \rbrace \subseteq \mathcal{X}, \\
   [\![ X,Y ]\!] & := \lbrace \left( X(\omega), Y(\omega) \right) : \omega \in \Omega \rbrace \subseteq \mathcal{X} \times \mathcal{Y}, \\
   [\![ X \vert y ]\!] & := \lbrace X(\omega): Y(\omega) = y, \omega \in \Omega \rbrace \subseteq \mathcal{X}.  
 \end{align}  
The dependence on $\Omega$ will normally be hidden, with most properties of interest expressed in terms of operations on these ranges.
As a convention, uv's are denoted by upper-case letters, while their realizations are indicated in lower-case.
The family $\left\{\llb X|y\rrb: y\in\llb Y\rrb\right\}$ of conditional ranges is denoted $\llb X|Y\rrb$.

\begin{definition}[Unrelatedness \cite{b9}  ]
\label{unrelatedDefn}
 The uvs $X_1, X_2, \cdots X_n $ are said to be \emph{(mutually) unrelated} if
 \beq
   [\![ X_{1}, X_{2}, \cdots, X_n ]\!] = [\![ X_{1}]\!] \times [\![ X_{2}]\!] \times \cdots \times [\![ X_{n}]\!].
 \eeq
\end{definition}

{\em Remark:} Unrelatedness, which is closely related to the notion of {\em qualitative independence} \cite{renyi70} between discrete sets, can be shown to be equivalent to the conditional range property
 \beq
   [\![ X_{k} | x_{1:k-1} \!] = [\![ X_{k}]\!],  \ \forall x_{1:k-1}\in   [\![ X_{1:k-1}]\!], \  k\in [2:n].
\label{unrelatedconditionalform}
 \eeq

\begin{definition}[Markovianity \cite{b9}]
The uvs $X_1, X_2$ and $Y$ are   said to form a {\em Markov uncertainty chain} $X_1\lra Y \lra X_2$ if 
\beq
[\![ X_1|y,x_2]\!] = [\![ X_1|y]\!], \ \ \forall (y,x_2)\in [\![ Y,X_2]\!].
\eeq
\end{definition}

{\em Remark:}  This can be shown to be equivalent to $X_1$ and $X_2$ being {\em conditionally unrelated given} $Y$, i.e.
\beq
   \llb X_{1}, X_{2}| y \rrb = 
\llb X_{1} | y \rrb \times \llb  X_{2} | y \rrb , \  \forall y\in \llb Y\rrb .
\label{condunrelated}
 \eeq
By the symmetry of \eqref{condunrelated}, $X_1 \lra Y\lra X_2$  iff $X_2 \lra Y \lra X_1$.

\subsection{System Model}

Consider a multiple access communication system with one receiver, two transmitters and three messages, as illustrated in Fig. (\ref{fig:DMAC}). 
Assume the messages $M^0$, $M^1$ and $M^2$ are mutually unrelated and finite-valued. 
Without loss of generality, for $i=0,1,2$ let $M^i$ take the integer values $[1:\mu^i]$ for some integer $\mu^i\geq 1$. For a given block-length $n\geq 1$, the messages are encoded into channel input sequences $X^1_{1:n}$ and $X^2_{1:n}$ as
\beq
X^j_{1:n}= \gamma^i (M^0,M^j), \ \  j=1,2,
\label{code}
\eeq
where $\gamma^1$ and $\gamma^2$ are the coding functions at each transmitter. Observe that the \emph{common message} $M^0$ is seen by both transmitters, while  the {\em private messages} $M^1$ and $M^2$ are available only to their respective transmitters.
The code rate for each message is defined as 
\beq
R^i := ( \log_2\mu^i)/n, \  \ i=0,1,2.
\label{ratedef}
\eeq
 Due to the common message, the two channel input sequences applied will typically be related. In the case where the common message can take only one value, so that $R^0=0$, each channel input is generated in isolation and is mutually unrelated with the other. At the other extreme, if the private messages  can each take only one value so that $R^1=R^2=0$, then the channel inputs are generated in complete cooperation. 

The encoded data sequences are then sent through a stationary memoryless MAC as depicted in Fig. (\ref{fig:DMAC}). 
The output $Y_k\in\mcY$ of the MAC is given in terms of a fixed function $f:\mcX_1\times\mcX_2\times\mcW\to\mcY$ as
\beq
   Y_{k} = f(X^1_{k},X^2_{k},W_k) \in\mcY,  \  \ k=1,2,\ldots,
\label{MAC}
\eeq
where $W_k$  is channel noise that is mutually unrelated with $W_{1:k-1}$, $M^0$, $M^1$, $M^2$, and has constant range $\llb W_k\rrb = \mcW$. 

At the receiver , the decoder $\delta$ produces message estimates $\hat{M}^0$, $\hat{M}^1$ and $\hat{M}^2$ from the channel output sequence $Y_{1:n}$. 
Under a zero-error objective, these estimates must always be exactly equal to the original messages, regardless of channel noise or interference between $X^1_{k}$ and $ X^2_{k}$. In other words, $\llb M^i|y_{1:n}\rrb$ is a singleton for each $i=0,1,2$ and any $y_{1:n}\in\llb Y_{1:n}\rrb$. For a given block-length $n$, we define the {\em zero-error $n$-capacity region} $\mcC_{0,n}$ of the MAC as the set of rate tuples $R = (R^i)_{i=0}^3$ for which this is possible by suitable choice of coding functions.

The system set-up above is inspired by that of \cite{b6}. The critical difference is that the messages and channel here are not assumed to have any statistical structure, and the aim is to recover the messages perfectly, not just with arbitrarily small error probability. In addition, we are interested in characterising the zero-error capacity region at finite $n$, not just as $n\to\infty$.


\tikzstyle{block} = [draw, fill=white, rectangle, minimum height=4em, minimum width=3em]
\tikzstyle{input} = [coordinate]
\tikzstyle{output} = [coordinate]
\tikzstyle{pinstyle} = [pin edge={to-,thin,black}]
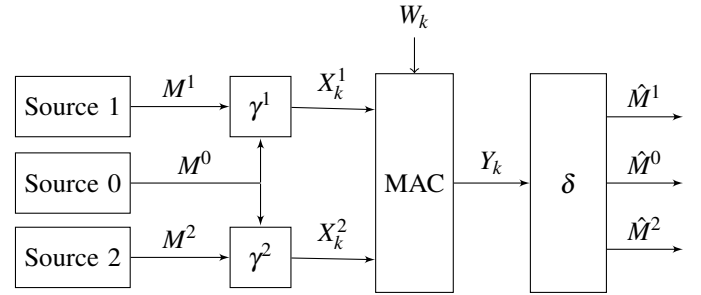
\begin{figure}[htbp]
 \centering
   \begin{tikzpicture}[auto, node distance=2cm,>=latex'] [scale=2]
    \node[block, name=E1] [minimum width=8mm, minimum height=8mm] (E1) {$\gamma^1$};      
    \node[block, below of= E1] [minimum width=8mm, minimum height=8mm] (E2) {$\gamma^2$}; 
    \node[block, left = 1.3cm of E1] [minimum width=12mm, minimum height=8mm] (Source1) {Source 1};
    \node[block, below =.2cm of Source1] [minimum width=12mm, minimum height=8mm] (Source0) {Source 0};
    \node[block, below of = Source1] [minimum width=12mm, minimum height=8mm] (Source2) {Source 2};
    \node[inner sep=0,minimum size=0, below = 0.6cm of E1] (cent) {};  
    \node[block, right = 1.5cm of cent] [minimum width=10mm, minimum height=29mm,pin={[pinstyle]above:$W_{k}$}] (MAC) {MAC};
    \node[block, right = 1cm of MAC] [minimum width=10mm, minimum height=29mm] (D) {$\delta$};
    \node[inner sep=0,minimum size=0, left of=E1] (k1) {};  
    \node[inner sep=0,minimum size=0, left of=E2] (k2) {};  
    \node[inner sep=0,minimum size=0, below =.6cm of E1] (k0) {};  
    \node[inner sep=0,minimum size=0, below =0cm of E1] (l1) {};   
    \node[inner sep=0,minimum size=0, above =0cm of E2] (l2) {};   
    \node[inner sep=0,minimum size=0, right of=E1] (d1) {}; 
    \node[inner sep=0,minimum size=0, right of=E2] (d2) {};
    \node [input, left = 1.3cm of E1] (input1) {};
    \node [input, below of= input1] (input2) {};
    \node [output, right = 1cm of D.-60] (output1) {};
    \node [output, right = 1cm of D] (output2) {};
    \node [output, right = 1cm of D.60] (output3) {};
    \draw [draw,->] (input1) -- node {$M^1$} (E1);
    \draw [draw,->] (input2) -- node {$M^2$} (E2);
    \draw[draw,-] (Source0) -- node {$M^0$} (k0);
    \draw[draw,->] (k0)   --(l1);
    \draw[draw,->] (k0)   -- (l2);
    \draw [draw,->] (E1)  -- node {$X^1_{k}$} (MAC.118);
    \draw [draw,->] (E2)  -- node {$X^2_{k}$} (MAC.243);
    \draw [draw,->] (MAC) -- node {$Y_{k}$} (D);
    \draw [->] (D.-60) -- node [name=Y] {$\hat{M}^2$}(output1);
    \draw [->] (D) -- node [name=Y] {$\hat{M}^0$}(output2);
    \draw [->] (D.60) -- node [name=Y] {$\hat{M}^1$}(output3);
   \end{tikzpicture}
    \caption{The two-transmitter MAC system with a common message operating at time instant $k$.}
    \label{fig:DMAC}
\end{figure}

\section{Nonstochastic Information and MAC Zero-Error Capacity}
\label{sec:ZECapacity}

In this section, we use the nonstochastic information concepts of \cite{b9,b1} to give an exact characterisation of the zero-error capacity region of the multiple access channel (MAC) defined in the previous section.

\subsection{Preliminaries on Nonstochastic Information}
First we present some necessary background concepts. Throughout this subsection $X$, $Y$,  $Z$, $Z'$ and $W$ denote  uncertain variables (uv's).

\begin{definition}[Overlap Connectedness \cite{b9}]
Two points $x$ and $x' \in [\![ X]\!] $ are said to be $[\![ X \vert Y ]\!]${\em -overlap connected}, denoted $x \leftrightsquigarrow  x'$, if there exists a finite sequence $\lbrace X \vert y_i \rbrace_{i=1}^{m} $ of conditional ranges such that $ x \in [\![ X \vert y_1 ]\!], x' \in [\![ X \vert y_m ]\!]$ and $[\![ X \vert y_i ]\!] \cap [\![ X \vert y_{i-1}]\!] \neq \emptyset$, for each $i \in \left[ 2, \cdots, m\right] $. 
\end{definition} 

{\em Remarks:} It is easy to see that overlap connectedness is both transitive and symmetric, i.e. it is  an equivalence relation between $x$ and $x'$. Thus it induces disjoint  {\em equivalence classes} that cover $\llb X\rrb$ and form a unique partition. This is called the 
 $[\![ X |Y ]\!]${\em -overlap partition}, denoted by $[\![ X \vert Y ]\!]_*$. 

\begin{definition}[Nonstochastic Information \cite{b9}]
The {\em nonstochastic information} between $X$ and $Y$ is given by
\beq
I_*[X;Y] = \log_2 \left | \llb X|Y\rrb_*\right |.
\label{defIstar}
\eeq
\end{definition} 

{\em Remark:} This can be shown to be symmetric, i.e. $I_*[X;Y] =  I_*[Y;X] $. 


\begin{definition}[Common Variables \cite{shannonLatticeTIT1953,wolfITW2004}]

A uv $Z$ is said to be a {\em common variable (cv)}  for $X$ and $Y$ if there exist  functions $f$ and $g$ such that
$Z = f(X) = g(Y)$. 

It is further said to be a {\em maximal} cv if any other cv $Z'$ admits a function $h$ such that $Z' = h(Z)$.

\end{definition}

{\em Remarks:} In the context of random variables, these concepts were first discussed by Shannon \cite{shannonLatticeTIT1953}, who used the term {\em common information element} for a maximal cv. Notice that no  cv can take more distinct values than the maximal one.

The nonstochastic information $I_*[X;Y]$ is precisely the log-cardinality of the range of a maximal cv between $X$ and $Y$. This is because it can be shown that $\forall (x,y) \in \llb X,Y\rrb$,  the partition set in $\llb X|Y\rrb_*$ that contains $x$ also uniquely specifies the set in $\llb Y|X\rrb_*$ that contains $y$. Thus these overlap partitions define a cv for $X$ and $Y$, with corresponding functions $f$ and $g$ given by the labelling. Furthermore, this cv can be proved to be maximal. See \cite{b1} for details.

\begin{definition}[Conditional $I_*$] The {\em conditional nonstochastic information} between  $X$ and $Y$ given  $W$  is 
\beq
I_*[X;Y|W]:= \min_{w\in\llb W\rrb } \log_2 \left | \llb X |Yw \rrb_* \right |,
\label{defcondinfo}
\eeq
where for a given $w\in\llb W\rrb$, $\llb X |Yw \rrb_*$ is the overlap partition of $\llb X|w\rrb$ induced by the family $\llb X |Yw \rrb$  of conditional ranges $\llb X |yw \rrb$, $y\in\llb Y|w\rrb$ \cite{b9}. 
\end{definition}

{\em Remark:}  It can be shown that $I_*[X;Y|W]$ also has an important interpretation in terms of cv's: it is the maximum log-cardinality of the ranges of all cv's $Z = f(X,W) = g(y,W)$ that are unrelated with $W$.
See \cite{b9} for details.

\subsection{MAC Zero-Error Capacity via Nonstochastic Information}

We are now in a position to prove the main result of this paper.

\begin{theorem}
For a given block-length $n\geq 1$, let  $\mathcal{R}(U,X^1_{1:n}, X^2_{1:n})$ be the set of nonegative tuples $(R_0,R_1,R_2)$ such that
\begin{align}
      n R_0 & \leq  I_{*}[U ;  Y_{1:n}]               \label{ineq:1}  \\
      n R_1 & \leq  I_{*}[X^1_{1:n};Y_{1:n} \vert U]  \label{ineq:2}  \\
      n R_2 & \leq  I_{*}[X^2_{1:n};Y_{1:n}\vert U]   \label{ineq:3}
\end{align}
where $X^i_{1:n}$, $i=1,2$, are sequences of inputs to the multiple access channel  (MAC) \eqref{MAC}, $Y_{1:n}$ is the corresponding channel output sequence, and $U$ is an auxiliary uncertain variable (uv). 

Then the {\em zero-error $n$-capacity region} $\mathcal{C}_{0,n}$ of the MAC over $n$ uses coincides with the union of the regions $\mathcal{R}(U,X^1_{1:n}, X^2_{1:n})$ over all uv's $U,X^1_{1:n}, X^2_{1:n}$ that satisfy the 
Markov uncertainty chains 
 $X^1_{1:n} \leftrightarrow U \leftrightarrow X^2_{1:n}$ and $ U \leftrightarrow \lp X^1_{1:n}, X^2_{1:n}\rp  \leftrightarrow Y_{1:n}$.

\label{conjecture:DMAC}
\end{theorem}   
 
{\em Remarks:} This result is the zero-error analogue of the Slepian-Wolf ordinary capacity region $\mcC$ \eqref{slepian-wolf}, in terms of nonstochastic rather than Shannon information. Although $\mcC$ is {\em prima facie} given in `single-letter' terms,  it is operationally relevant only at large block-lengths $n$, to yield small probabilities of error. In contrast, the result above specifies all rates tuples that allow {\em exactly} zero errors to be achieved at a given finite $n$.  This could potentially be of interest in safety-critical, low-latency  applications in distributed networked control.
If arbitrarily long blocks are permitted,  then the relevant zero-error capacity region $\mcC_0$ is given by the convex hull of $\cup_{n\geq 1} \mathcal{C}_{0,n}$. 

Although \eqref{ineq:1}--\eqref{ineq:3} give a cuboidal rate region $\mathcal{R}(U,X^1_{1:n}, X^2_{1:n})$ , it is not clear  if the zero-error capacity regions also have geometrically simple shapes, due to the unions over $U,X^1_{1:n}, X^2_{1:n}$ and  $n$. We aim to investigate this in future work, for specific channels of interest.
\subsubsection{Proof of Converse}

Consider a zero-error code \eqref{code} with block-length $n$ operating at rates $R_0, R_1$ and $R_2$ \eqref{ratedef} over the MAC \eqref{MAC}, and set $U=M^0$.  As $M^i$, $i=0,1,2$ are mutually unrelated, it follows from \eqref{code} that the codewords $X^1_{1:n}$ and $X^2_{1:n}$ are conditionally unrelated given $M^0$,
i.e. the first Markov uncertainty chain $X^1_{1:n} \leftrightarrow U \leftrightarrow X^2_{1:n}$ is satisfied. Since the channel noise in \eqref{MAC} is unrelated with the messages and hence with the codewords, we also have the second Markov uncertainty chain $Y_{1:n} \lra \lp X^1_{1:n}, X^2_{1:n}\rp \lra U$.

As the messages are all errorlessly recovered at the receiver, there certainly exists a decoding function $\delta^0$ such that $M^0 = \delta^0(Y_{1:n})$. 
Setting $U=M^0$, we see that $M^0$ is therefore a common variable (cv) between $U$ and $Y_{1:n}$. By the maximal cv property of $I_*$,
\beq
nR^0\equiv \log_2|\llb M^0\rrb| \leq I_*[U; Y_{1:n}],
\eeq
proving \eqref{ineq:1}.
%

We next prove the remaining two inequalities. Observe that for a given realisation  $m^0$ of the common message,
there must be a unique message $m^1$ corresponding to each channel codeword $x^1_{1:n}$; otherwise, multiple values of $m^1$ would be associated with a single channel output sequence $y_{1:n}$, violating the zero-error requirement.
Consequently, there must exist a mapping $g$ such that $M^1 = g(X^1_{1:n}, M^0)$. Furthermore, by the zero-error property there also exists a function $\delta^1$ such that
$M^1 = \delta^1(Y_{1:n})$. 

Thus $M^1$ is a cv between $(X^1_{1:n}, M^0)$ and $(Y_{1:n}, M^0)$. As by hypothesis it is also unrelated with $U=M^0$, the interpretation of conditional $I_*$ in terms of maximal unrelated cv's allows us to conclude that
\beq
nR^1\equiv \log_2|\llb M^1\rrb| \leq I_*[X^1_{1:n}; Y_{1:n}|M^0]= I_*[X^1_{1:n}; Y_{1:n}|U],
\eeq
proving  \eqref{ineq:1}.  In a similar way, the bound on the rate $R^2$ stated in (\ref{ineq:3}) can be shown.


\subsubsection{Proof of Achievability}
We now prove that if we have a block-length $n$ and uv's $U$, $X_{1:n}^j$, $j=1,2$ satisfying the requirements in 
Theorem \ref{conjecture:DMAC}), it is possible to construct a zero-error coding scheme at rates achieving equality in \eqref{ineq:1}--\eqref{ineq:3}.

\paragraph{Codebook Generation}
First, set $nR^0 = I_*[ U; Y_{1:n}]$ and pick one point in each of the disjoint sets of the overlap partition $[\![ U \vert  Y_{1:n}]\!]_* $.
With mild abuse of notation call these distinct points $u(m^0)$, $m^0 = 1,\ldots , 2^{nR_0}$.

\begin{figure}
\centering
  \begin{tikzpicture}[scale=.95]
    \draw[-latex] (0,0) -- (8,0) node [right] {$U$};
    \draw (0,-.2) -- (0, .2);
    \draw[latex-latex] (0,.1) -- (1.2,.1);
    \node at (0.6,0) [below] {\small $u_1$};
    \fill[red] (0.6,0) circle (1.5pt);
    \draw (1.2,-.2) -- (1.2, .2);
    \draw[latex-latex] (1.2,.1) -- (2.6,.1);
     \node at (1.9,0) [below] {\small $u_2$};
     \fill[red] (1.9,0) circle (1.5pt);
    \draw (2.6,-.2) -- (2.6, .2);
    \draw[latex-latex] (2.6,.1) -- (4,.1);
    \node at (3.3,0) [below] {\small $u_3$};
    \fill[red] (3.3,0) circle (1.5pt);
    \draw (4,-.2) -- (4, .2);
    \draw[latex-latex] (4,.1) -- (6,.1);
    \node at (5,0) [above] {\small $\cdots$};
    \node at (5,0) [below] {\small $ \cdots$};
    \draw (6,-.2) -- (6, .2);
    \draw[latex-latex] (6,.1) -- (7.5,.1);
    \node at (6.75,0) [below] {\small $u_{2^{nR^0}}$};
    \fill[red] (6.75,0) circle (1.5pt);
    \draw (7.5,-.2) -- (7.5, .2);
  \end{tikzpicture}
  \caption{Example of an overlap partition $[\![ U \vert Y_{1:n} ]\!]_*$. The horizontal lines represent to the different member-sets of each partition and the filled circles correspond to the selected points $u_i$.}
  \label{fig:overlapPartitionU}
\end{figure}
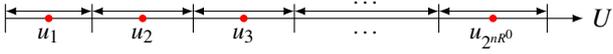

Next, observe that since $nR^i = I_*[X^i_{1:n};Y_{1:n}|U]$ for $i=1,2$, \eqref{defcondinfo} implies that
\beq
2^{nR^i} \leq \left | [\![ X^i_{1:n} | Y_{1:n}, U=u(m^0) ]\!]_* \right |, \  i=1,2, m^0 \in [1:2^{nR^0}].
\label{RiIneq}
\eeq
For any $m^0$, we may therefore pick $2^{nR^i}$ distinct codewords $x^i_{1:n}$ from $[\![ X^i_{1:n} | U=u(m^0) ]\!]$ such that there is at most one codeword in each set of the overlap partition
$[\![ X^i_{1:n} | Y_{1:n}, U=u(m^0) ]\!]_*$. 
Denote  these codewords by $\gamma^i(m^0,m^i)$, $m^i\in[1:2^{nR^i}]$.
This gives us our coding laws \eqref{code}.

\paragraph{Zero Error}

To show that this code may be decoded with zero error, observe first that 
since $X^1_{1:n} \leftrightarrow U \leftrightarrow X^2_{1:n}$, the joint conditional range
$[\![ X^1_{1:n},  X^2_{1:n}|  U=u(m^0) ]\!]$ is just the Cartesian product
\[
 [\![ X^1_{1:n} | ,U=u(m^0) ]\!] \times [\![ X^2_{1:n} |  U=u(m^0) ]\!].
\]
Thus we are guaranteed that for every $m^0$, all codeword pairs $\lp \gamma^1(m^0,m^1), \gamma^2(m^0,m^2)\rp$, 
$m^i=1,\ldots, 2^{nR^i}$, $i=1,2$,  lie within the conditional joint range
$[\![ X^1_{1:n},  X^2_{1:n}|  U=u(m^0) ]\!]$. In other words, for every combination of $m^0, m^1$ and $ m^2$, the triple
$\lp \gamma^1(m^0,m^1), \gamma^2(m^0,m^2),  u(m^0) \rp $ is a valid point inside the joint range 
$\llb X^1_{1:n},  X^2_{1:n}, U\rrb$. 

The decoding proceeds in three stages. In the first stage, the common message $m^0$ is recovered.
Recall that each of the $2^{nR^0}$ points $u(m^0)$ lies in a distinct set of the overlap partition $\llb U| Y_{1:n}\rrb_*$.
By the common variable (cv) property of overlap partitions, this set is uniquely determined by the corresponding set of the matching overlap partition 
$\llb  Y_{1:n}| U\rrb_*$ that contains the channel output sequence $y_{1:n}$. In this way, $m^0$ is uniquely decoded.

In the second stage, having recovered $m^0$, the decoder calculates which distinct set of the conditional overlap partition $[\![ Y_{1:n} | X^1_{1:n}, U=u(m^0) ]\!]_*$ contains $y_{1:n}$. Again by the cv property, this set uniquely determines the corresponding set of the matching conditional overlap partition $[\![ X^1_{1:n} | Y_{1:n}, U=u(m^0) ]\!]_*$ that contains the codeword $\gamma^1(m^0,m^1)$. By construction, for each $m^0$ there is at most one codeword in each set of this latter conditional overlap partition; thus $m^1$ is uniquely recovered.  

In the third stage, the decoder repeats the second stage but with $X^2_{1:n}$ instead of $X^1_{1:n}$, and recovers $m^2$ uniquely in the same way.

\section{Noncooperative Connectedness and Information}
\label{sec:NCConnectedness}

The notions of overlap connectedness and common variables (cv's) were critical in developing a characterisation of MAC zero-error capacity based on nonstochastic information. In this section, we consider a related but more basic problem, in which three uncertain variables $X_1,X_2$ and $Y$ with joint range $\llb X_1,X_2,Y\rrb$ are respectively observed by three agents. The agents observing $X_1$ and $X_2$ each wish to separately deduce as much as possible about $Y$, while the agent observing $Y$ wishes to know exactly what the other two agents have deduced about it. In other words, we seek to characterise cv's of the form
\beq
Z= \lp f_1(X_1) \ f_2(X_2) \rp = \lp g_1(Y) \ g_2(Y) \rp \equiv g(Y)
\label{nonco}
\eeq
In order to do so, we propose a new notion of connectedness and nonstochastic information.


\begin{definition} \emph{(NC-Connectedness)}
A pair of points $(x_1,x_2,y) $ and $(x_1',x_2',y') \in [\![ X_1, X_2,Y]\!]$ is called \emph{noncooperatively (NC-)connected}, denoted $(x_1,x_2,y) \overset{NC}{\leftrightsquigarrow} (x_1',x_2',y')$, if
\begin{enumerate}[label={(\roman*)}]
  \item $x_1 \leftrightsquigarrow x_1'$ in $[\![ X_1 \vert Y]\!]$,\\ 
and 
  \item $x_2 \leftrightsquigarrow x_2'$ in $[\![ X_2 \vert Y]\!]$, 
\end{enumerate} 
where the symbol "$\leftrightsquigarrow$" refers to overlap connectedness. 
\label{def:MACOverlapConnectivity}
\end{definition}

{\em Remark:} It is clear that NC-connectedness inherits the symmetry and transitivity of overlap connectedness; thus it is an equivalence relation, which splits
$[\![ X_1, X_2,Y]\!]$ into disjoint equivalence classes. Call this partition the {\em NC-partition} of $\llb X_1,X_2,Y\rrb$.

From the definition, it can be shown that each set of the NC-partition is uniquely defined  by a set in the product $\llb X_1| Y\rrb_*\times \llb X_2|Y\rrb_*$ of overlap partitions.
By the common variable property of overlap partitions, it is also uniquely defined by a corresponding pair of sets in 
the matching overlap partitions $\llb Y| X_1\rrb_*$ and $\llb Y|X_2\rrb_*$. As both these latter partitions are of $\llb Y\rrb$, this pair of sets is uniquely defined by a more refined set in the pairwise intersection or {\em join} $\llb Y| X_1\rrb_*\vee \llb Y|X_2\rrb_*$.

Thus the overlap partitions $\llb X_1| Y\rrb_*$ and $\llb X_2|Y\rrb_*$ yield the functions $f_1(X_1)$ and $ f_2(X_2)$ of \eqref{nonco}, while $\llb Y| X_1\rrb_*\vee \llb Y|X_2\rrb_*$ yields the matching function $g(Y)$.

\subsection{Maximal Common Variable and Noncooperative $I_*$}
\begin{theorem}
The functions $f_1$, $f_2$ and $g$ given respectively by the (labels of) the partitions $\llb X_1|Y\rrb_*$, $\llb X_2|Y\rrb_*$ and $\llb Y| X_1\rrb_*\vee \llb Y|X_2\rrb_*$ yield a common variable (cv)  $Z_*$ in the sense \eqref{nonco} that is maximal. That is, any other cv 
\[
Z=  \lp \bar{f}_1(X_1), \bar{f}_2(X_2) \rp =\bar{g}(Y)
\] 
 admits a function $h$ such that $Z=h(Z_*)$.
\end{theorem}


\textit{Proof.} This statement can be proven by contradiction. Suppose that there is a set $\mcP$ in the NC-partition of $\llb X_1,X_2,Y\rrb$ that is not wholly contained inside any partition set induced by the cv $Z$.
Then there must exist two admissible points  $(x_1,x_2,y)$ and $(x_1', x_2',y')$ in $\mcP$ that lie in different partition sets of $Z$, therefore yielding different values $z\neq z'$ of $Z$.  That is,
\beq
   \lp \bar{f}_1(x_1), \bar{f}_2(x_2) \rp \neq 
 \lp \bar{f}_1(x_1'), \bar{f}_2(x_2') \rp
\label{eq:contradiction}   
\eeq

\begin{figure}
 \centering
 \begin{tikzpicture}
  \draw [fill=gray!20, rotate=0] (3,2) ellipse (2.7 and 1);
  \draw [thick,-latex] (0,0) -- (6,0) node [right] {$$};
  \draw [thick,-latex] (0,0) -- (0,4) node [above] {$$};
  \draw [dashed] (4.5,1) -- (1,2.8) node [above] {{\small \emph{$Z$-Partition}}};   
  \coordinate (X) at (1.5,1.5);
  \fill (X) circle (1.5pt) node [above] {$(x_1,x_2,y)$};
  \coordinate (X2) at (4,2.5);
  \fill (X2) circle (1.5pt) node [below] {$(x_1',x_2',y')$};
  \node (X3) at (3.2,.7) {{\small \emph{Equivalence Class under $NC-$Connectedness}}};
\end{tikzpicture}
\caption{Illustration of the scenario expressed by (\ref{eq:contradiction}).}
  \label{fig:example1}
\end{figure}
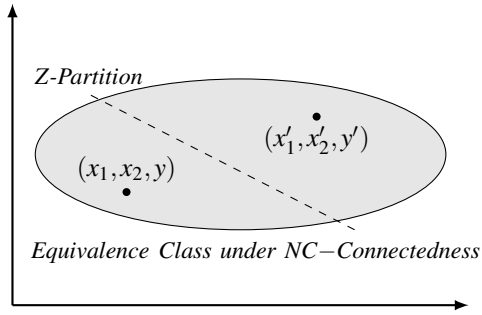 

Without loss of generality, say that $\bar{f}_1(x_1) \neq \bar{f}_1(x_1')$. 
As $Z$ is a cv in the sense \eqref{nonco}, its first component is  a cv $Z_1 = \bar{f}_1(X_1) = \bar{g}_1(Y)$ between $X_1$ and $Y$. 

However,
by the maximal cv property,  $Z_1$ is a function of the set of  the overlap partition $\llb X_1|Y\rrb_*$ that contains $X_1$.
Since both $x_1$ and $x_1'$ lie in the same set $\llb X_1|Y\rrb_*$, they must therefore yield the same value $\bar{f}_1(x_1) = \bar{f}_1(x_1')$,
contradicting \eqref{eq:contradiction}.
%
%
%

With this result, it is  then natural to take the log-cardinality of $\llb Z_*\rrb$ as a new measure of nonstochastic information,

\beq
I_*^{\mathrm{NC}}[X_1,X_2;Y] := \log_2 \left | \llb Y| X_1\rrb_*\vee \llb Y|X_2\rrb_*\right |.
\label{defNCinfo}
\eeq

\section{Conclusion}
\label{sec:conclusion}
In this paper, zero-error multiple access communication systems was analysed in a nonprobabilistic framework using uv's and nonstochastic information. These notions were used to characterise the zero-error capacity region of multiple access channels. The presented analysis is not only valid for asymptotically large blocklength but it also includes the case of finite $n$. Subsequently, theconcept of noncooperative connectedness was introduced and used as a tool to extend the concept of nonstochastic information to non-cooperative situations.

Future work will consider the extension of this framework to include the general multi-user case (more than two input sequences) and MAC's with feedback. These scenarios represent the first steps of modelling information flows in distributed estimation and control systems using  nonstochastic concepts.

\end{document}